# Delta-doped β-Ga$_2$O$_3$ thin films and β-(Al$_{0.26}$Ga$_{0.74}$)$_2$O$_3$/β-Ga$_2$O$_3$ heterostructures grown by metalorganic vapor-phase epitaxy

Praneeth Ranga[1]*, Arkka Bhattacharyya[1], Ashwin Rishinaramangalam[2], Yu Kee Ooi[1], Michael A. Scarpulla[1,3], Daniel Feezell[2], Sriram Krishnamoorthy[1]*

[1]Department of Electrical and Computer Engineering, The University of Utah, Salt Lake City, UT 84112, United States of America

[2] Center for High Technology Materials, University of New Mexico, Albuquerque, NM 87106, United States of America

[3] Department of Materials Science and Engineering, The University of Utah, Salt Lake City, UT, 84112, United States of America

E-mail: praneeth.ranga@utah.edu, sriram.krishnamoorthy@utah.edu

We report on silicon delta doping of metalorganic vapor-phase epitaxy grown β-Ga$_2$O$_3$ thin films using silane as precursor. Delta-doped β-Ga$_2$O$_3$ epitaxial films are characterized using capacitance-voltage profiling and secondary-ion mass spectroscopy. The sheet charge density is in the range of 2.9 x 10$^{12}$ cm$^{-2}$ to 9 x 10$^{12}$ cm$^{-2}$ with a HWHM (towards substrate) ranging from 6.2 nm – 3.5 nm is measured. We also demonstrate a high-density (n$_s$ - 6.4 x 10$^{12}$ cm$^{-2}$) degenerate electron sheet charge in a delta-doped β-(Al$_{0.26}$Ga$_{0.74}$)$_2$O$_3$/β-Ga$_2$O$_3$ heterostructure. The total charge could also include a contribution from a parallel channel in the β-(Al$_{0.26}$Ga$_{0.74}$)$_2$O$_3$ alloy barrier.

β-Ga$_2$O$_3$ is an ultra-wide bandgap semiconductor (E$_g$~ 4.6 eV) with predicted high critical breakdown field strength (6 – 8 MV/cm), making it an attractive alternative to other wide bandgap semiconductors[1]. Very high electron mobility values close to theoretically predicted[2–4] maximum of 200 cm$^2$/V.s have been reported in β-Ga$_2$O$_3$ thin films grown by metalorganic vapor-phase epitaxy[5–7] (MOVPE) and hydride vapor-phase epitaxy[8] (HVPE). Critical breakdown field strength exceeding GaN has already been demonstrated by multiple groups in both vertical[9,10] and lateral power devices[11]. In addition to high-power applications, β-Ga$_2$O$_3$ is also attractive for high-frequency applications[12] such as RF power amplifiers and RF switches[13], due to the combination of high breakdown field along with a saturation velocity of ~ 10$^7$ cm/s[14,15].

A variety of lateral transistor designs have already been studied using a uniformly doped β-Ga$_2$O$_3$ channel[16,17] and exfoliated β-Ga$_2$O$_3$[18–20]. The performance of these devices is limited mainly by the epitaxial film quality, dielectric interface and contact resistance. In contrast to uniformly-doped thin films, channels with extremely narrow dopant profiles (delta-doped channels) offer an alternative device topology, which can potentially enable superior performance than conventional MESFETs. In a delta-doped structure[21], the dopants are confined to a few monolayers thick semiconductor layers, leading to confinement of charge carriers by a potential well. The higher mobility observed in the delta-doped channel compared to a uniformly-doped channel with equivalent sheet charge can be attributed to spread of the electron wave function to the undoped regions on either side of the ionized donor charges, resulting in reduced scattering of the electron gas with by the ionized donors[22]. Delta-doped channels with silicon donors grown using molecular beam epitaxy have been used to realize a wide range of 2D electron sheet charge in GaAs, from 2 x 10$^{12}$ cm$^{-2}$ to 8 x 10$^{12}$ cm$^{-2}$ [23]. The high 2D sheet charge in combination with a short gate to channel distance can lead to higher transconductance[13,24,25] in delta-doped FETs than conventional MESFETs. The peak electric field in a delta-doped MESFET is lower than the peak field value of a conventional MESFET with a uniformly doped channel (with identical gate to channel distance and equivalent sheet charge)[26]. Consequently, delta-doped channels are expected to exhibit superior breakdown characteristics. Delta-doped devices can also be scaled to shorter device dimensions maintaining the required aspect ratio for good electrostatic control and do not suffer from significant short channel effects. This enables aggressive scaling of the gate length for high-frequency operation. By utilizing a short gate length of 120 nm on a delta-doped channel of ~1 x 10$^{13}$ cm$^{-2}$ charge, a high drain current of 260 mA/mm and f$_T$ of 27 GHz have already been demonstrated in β-Ga$_2$O$_3$[27].

Delta-doped channels still suffer from significant ionized impurity scattering, due to the overlap and close proximity of ionized donor atoms to the electron sheet charge. In a modulation-doped channel, the ionized donors and the 2DEG (two-dimensional electron gas) are spatially separated, this results in considerable reduction of ionized impurity scattering in the channel. The study of delta doping in the β-(Al$_x$Ga$_{1-x}$)$_2$O$_3$ barrier layer is also crucial for obtaining a high mobility 2DEG at the β-(Al$_x$Ga$_{1-x}$)$_2$O$_3$/β-Ga$_2$O$_3$ heterostructure interface. Based on transport theory, when 2DEG density exceeds 5 x 10$^{12}$ cm$^{-2}$, electron mobility is expected to be much larger than that of bulk β-Ga$_2$O$_3$ due to enhanced screening of phonon modes[28]. 2DEG sheet charge of 2 x 10$^{12}$ cm$^{-2}$ with an electron mobility of 180 cm$^2$/V.s has been reported in MBE-grown β-(Al$_x$Ga$_{1-x}$)$_2$O$_3$/ β-Ga$_2$O$_3$ MODFET. This is currently limited by the conduction band offset at the β-(Al$_x$Ga$_{1-x}$)$_2$O$_3$/ β-Ga$_2$O$_3$ heterojunction[29]. Using an ultra-thin spacer layer, 2DEG density as high as 6.1 x 10$^{12}$ cm$^{-2}$ and a channel mobility of 147 cm$^2$/V.s has been recently reported in a β−(Al$_{0.18}$Ga$_{0.82}$)$_2$O$_3$/ Ga$_2$O$_3$ heterosture [30].

Modulation-doping has been explored previously using molecular beam epitaxy[22,31,32], but efforts on modulation-doped MOVPE-grown heterostructures are still in early stages[33]. Previously, we demonstrated modulation doping using MOVPE-grown uniformly-doped β-(Al$_{0.26}$Ga$_{0.74}$)$_2$O$_3$ barrier layer[34] with a total electron sheet charge of 2.3 x 10$^{12}$ cm$^{-2}$. Based on the lever rule for charge control in modulation-doped heterostructures, the 2DEG charge density is maximized when the donor delta sheet is very close to the β-(Al$_x$Ga$_{1-x}$)$_2$O$_3$/β-Ga$_2$O$_3$ heterointerface[30]. For obtaining a high-density 2DEG sheet charge with high mobility without a parallel channel in the alloy barrier, a heavily-doped delta sheet with an abrupt doping profile is required in conjunction with a high Al composition β-(Al$_x$Ga$_{1-x}$)$_2$O$_3$ barrier layer. In this work, we report on the study of delta doping in (010)-oriented β-Ga$_2$O$_3$ thin films grown by MOVPE, characterized using capacitance-voltage (CV) profiling and secondary-ion mass spectroscopy (SIMS) analysis. We also report modulation-doping of β-(Al$_{0.26}$Ga$_{0.74}$)$_2$O$_3$/β-Ga$_2$O$_3$ heterojunction, using a delta-doped β-(Al$_{0.26}$Ga$_{0.74}$)$_2$O$_3$ barrier. Using low-temperature CV measurements, we confirm the degenerate nature of the electron gas with a sheet charge of 6.4 x 10$^{12}$ cm$^{-2}$ at room temperature. The measured charge could also include a contribution from a parallel channel in the β-(Al$_{0.26}$Ga$_{0.74}$)$_2$O$_3$ alloy barrier.

Growth of β-Ga$_2$O$_3$ is performed using a MOVPE reactor (Agnitron Agilis) utilizing TEGa (Triethyl Gallium), O$_2$ and diluted silane (40 ppm silane diluted with Ar) as precursors and argon as the carrier gas. Growth is carried out at a temperature of 810 °C and a pressure of 15 Torr on (010)-oriented Sn-doped β-Ga$_2$O$_3$ bulk substrates (Novel Crystal Technology). The precursor flow values are set as follows: O$_2$ – 500 sccm, TEGa – (22-65) sccm and Argon – 1100 sccm. The samples are cleaned with solvents (Acetone, Methanol and DI water) followed by Piranha treatment before loading into the chamber. The epitaxial structure consists of an undoped cap layer along with a Si delta-doped layer grown on top of an unintentionally doped (UID) buffer layer. We used a growth interruption step for delta doping, similar to the process used in MOVPE growth of other delta-doped compound semiconductors[35]. First, a thick buffer of undoped β-Ga$_2$O$_3$ is grown at a constant growth rate (~ 7.5 nm/min). Once the desired buffer thickness is reached, the supply of TEGa is interrupted while keeping the argon (1100 sccm) and oxygen flow (500 sccm) the same. The chamber is then purged for 45 seconds to remove any unreacted TEGa source molecules. Next, diluted silane is supplied at a constant flow rate for 60 seconds under a constant oxygen flow (500 sccm). A post-purge step is carried out for 45 seconds to remove any unreacted silane from the chamber. Following the post-purge step, TEGa flow is resumed and continued until the desired cap layer thickness is reached. A series of samples are grown under different silane flows to understand silicon incorporation and the charge distribution. The purge times (45 seconds) and silane flow time (60 seconds) are kept constant for all the samples.

The list of samples used in this study and the corresponding epitaxial structure details (buffer layer thickness, cap layer thickness, silane flow used for delta doping) are summarized in Table I. Silane flows are increased from 1.9 to 17.3 nmol/min in samples A-D, with an aim to study delta doping over a wide range of sheet charge density. When the delta sheet charge density is low, we expect significant depletion of the sheet charge due to the Schottky barrier. To avoid significant surface depletion, a 90 nm thick cap layer is utilized for samples A and B. It is essential to completely deplete the delta-doped channel to extract the charge profile of the delta-doped layer from CV measurements. While the thick cap layer is sufficient to modulate and deplete the low delta sheet charge density, high-density channels could not be depleted before the onset of reverse leakage, due to the low capacitance resulting from a 90 nm thick cap layer. To circumvent this issue, we employed thinner cap layers (<35 nm) to study high-density delta-doped channels (Samples C and D). Additionally, a 35 nm thick $SiO_2$ dielectric layer is deposited on top of Sample D, to profile the forward tail of the delta sheet. Ti/Au (50/50 nm) metal stack deposited by DC sputtering is used to form ohmic contacts. Ni/Au (50/50 nm) stack deposited by ebeam evaporation is used to form Schottky contacts. CV measurements are performed on Schottky diodes using Keithley 4200 parametric analyzer to characterize the sheet charge density and spread of the charge distribution.

Figure 1 shows the depth profile of delta-doped samples extracted from CV measurements. CV measurements show the sheet charge density increasing with increase in silane flow, as expected. The peak carrier concentration changed from $1.4 \times 10^{18}$ $cm^{-3}$ (Sample A) to $5.4 \times 10^{18}$ $cm^{-3}$ (Sample D) with increasing silane flow. Total electron sheet charge is measured to be $2.9 \times 10^{12}$ $cm^{-2}$ and $6.2 \times 10^{12}$ $cm^{-2}$ in samples A and B, respectively. A sheet charge density of $9 \times 10^{12}$ $cm^{-2}$ and $7.8 \times 10^{12}$ $cm^{-2}$ is measured for samples C and D. We extracted a carrier concentration FWHM of 15.6 nm, 14.2 nm, 13.9 nm, 12.2 nm for samples A –D from the CV depth profile. HWHM of the charge profile (towards the substrate) is measured to be 6.2 nm, 4.7 nm, 3.6 nm and 3.5 nm for samples A, B, C and D respectively. A general trend of reducing HWHM is observed with increasing silane flow (see supplementary data). This is attributed to enhanced confinement of the wave function at high doping levels due to the deepening of the potential well. We also observe asymmetrical broadening of charge profile in samples A -D with the HWHM much larger towards the growth surface.

Multiple delta sheets with identical growth conditions as that of Samples A-C are grown on a Fe-doped substrate to understand silicon incorporation and spread of Si-donors in delta-doped β-$Ga_2O_3$. Layer E is grown under a silane flow of 34.7 nmol/min, which is twice the value of silane flow used for layer D. In addition to changing the silane flow (A-E), we also studied silicon incorporation at two different growth temperatures, namely 775 °C (layer G) and 845 °C (layer F), while keeping the same silane flow (34.7 nmol/min). Secondary-ion Mass Spectroscopy (SIMS) is performed (Eurofins EAG) on the stack to characterize the density of silicon donors and other impurities in the epilayers. To ensure that the interaction between two neighboring delta sheets is negligible, we included a thick UID layer(~150nm) between two adjacent delta sheets. The stack along with the SIMS scan for Si and H species is shown in Fig. 2.

The total Si atom density changes from $3.4 \times 10^{12}$ atoms.$cm^{-2}$ to $3.9 \times 10^{13}$ atoms.$cm^{-2}$ for silane flow ranging from 1.9 to 34.7 nmol/min. The net silicon concentration extracted from SIMS is observed to increase proportionally with respect to the silane flow (slope ~ 1). The delta sheet grown at 845 °C (layer F- $3.7 \times 10^{13}$ atoms.$cm^{-2}$) has a slightly lower charge density compared to layers grown at 810 °C (layer E- $3.9 \times 10^{13}$ atoms.$cm^{-2}$) and 775 °C (layer G- $3.9 \times 10^{13}$ atoms.$cm^{-2}$). This change can be attributed to the increased desorption of Si from the growth surface at elevated growth temperatures. Carbon concentration in our films is close to the SIMS detection limit. Interestingly, we observe a spike in hydrogen which tracks closely the Si-delta sheet, showing enhanced hydrogen incorporation at higher silane flows. In three of the delta-doped layers (layers E, F, and G) a significant amount of hydrogen incorporation is observed (total hydrogen concentration between $7 \times 10^{12}$ atoms.$cm^{-2}$ – $1.3 \times 10^{13}$ atoms.$cm^{-2}$). We also observe that the layer with lower growth temperature has the highest amount of H incorporation[36]. During the growth interruption step in the absence of TEGa, decomposition of silane results in $H_2$ by-product formation. If the $H_2$ is not purged away quickly, $H_2$ can get incorporated into the epitaxial film. Hydrogen concentration in the UID layer is close to the SIMS detection limit, strongly indicating that silane is the source of $H_2$ in our epilayers during growth interruption for delta doping.

To understand the electrical activity of Si donors in delta-doped layers, we compare the Si atom density from SIMS and measured sheet charge from the CV profile. Figure 3 shows both the SIMS and CV extracted concentration for delta sheets grown under identical conditions. At lower sheet charge density (Samples A-C), the Si atom density and electron density extracted from CV match very well, indicating close to 100 % activation of Si donors. At high Si atomic density (samples D) we see a large deviation of extracted sheet charge (CV) from the estimated silicon concentration. This suggests a strong compensation effect or passivation of donors[23], requiring further detailed transport measurements and annealing experiments to understand the discrepancy between SIMS and CV measurements.[37] The FWHM of the delta sheet layers in our work is significantly larger than the theoretically expected spread of the electron wave function (see supplementary information). Detailed growth optimization is required to develop strategies for a sharp delta doping profile in MOVPE-grown thin films.

We also utilized delta-doping in the $\beta$-$(Al_xGa_{1-x})_2O_3$ barrier to realize a high electron sheet charge at the $\beta$-$(Al_xGa_{1-x})_2O_3$/ $\beta$-$Ga_2O_3$ heterointerface. The epitaxial structure is shown in Fig. 4(a). The stack consists of 260 nm UID $\beta$-$Ga_2O_3$ buffer layer followed by 18 nm thick $\beta$-$(Al_xGa_{1-x})_2O_3$. Modulation doping is achieved by delta doping of $\beta$-$(Al_xGa_{1-x})_2O_3$ following the 1.3 nm thick undoped spacer layer growth. We used growth conditions identical to layer E with the growth being performed with the same silane molar flow (34.7 nmol/min), silane flow time (1 min), purge time (45 s) and growth temperature (810 °C). Growth is performed on a Sn-doped (010) $\beta$-$Ga_2O_3$ substrate to enable ohmic contact formation to the electron sheet charge. Aluminum barrier composition[38] of 26 % and $\beta$-$(Al_{0.26}Ga_{0.74})_2O_3$ barrier layer thickness of 18 nm is extracted from HRXRD (Panalytical Empyrean) 2$\theta$-$\omega$ scan, assuming complete strain. The measured Al barrier composition is very close to the [TMAl]/[TMAl+TEGa] molar flow ratio (25%), indicating no significant prereactions. Ti/Au (50/50 nm) stack is deposited using DC sputtering as the ohmic contact on the back surface of the wafer, followed by ebeam evaporated Ni/Au (50/50 nm) on the epilayer surface for Schottky contacts. The sample showed an RMS surface roughness of 2.2 nm (see supplementary data).

Room temperature CV measurements (Fig. 4 (c)) are performed on 300 μm diameter size circular Schottky pads. Total electron sheet charge of $6.4 \times 10^{12}$ cm$^{-2}$ is extracted from the depth profile (Fig. 4 (d)). In order to confirm the degenerate nature of the electron sheet, we performed low-temperature CV measurements (90 K) to observe any change in the total electron concentration. A sheet charge density of $6.1 \times 10^{12}$ cm$^{-2}$ is measured at 90 K, indicating a degenerate electron sheet charge. Considering that delta sheets in β-Ga$_2$O$_3$ have relatively large FWHM, we cannot rule out incorporation of Si-donors into the UID β-Ga$_2$O$_3$ channel. A considerable amount of charge could result from spreading of Si-donors in the β-Ga$_2$O$_3$ channel, in addition to modulation doped carriers from delta-doped barrier layer. In order to characterize electron mobility of the channel, the epitaxial structure needs to be grown on insulating substrates and direct ohmic contacts to the channel and ohmic contact regrowth are required[29]. Forward bias CV measurements indicate that the peak of the delta sheet is located in the β-(Al$_{0.26}$Ga$_{0.74}$)$_2$O$_3$ layer (see supplementary data). By increasing the spacer layer thickness to 4 nm and maintaining the total β-(Al$_{0.26}$Ga$_{0.74}$)$_2$O$_3$ thickness to be 18 nm, we observed a reduction in the electron charge to $2.5 \times 10^{12}$ cm$^{-2}$ (see supplementary data). A thicker spacer layer will increase the probablility of parallel channel formation in the β-(Al$_{0.26}$Ga$_{0.74}$)$_2$O$_3$ layer. However, the peak of the charge profile is measured to be at the heterojunction, indicating the absence of a parallel channel in the alloy barrier in the sample with 4 nm spacer. Low-temperature Hall measurements[29] are required to confirm the absence of a parallel channel in the β-(Al$_{0.26}$Ga$_{0.74}$)$_2$O$_3$ cap layer. Nevertheless, the high-density degenerate electron gas demonstrated in this work establishes the promise of MOVPE growth technique to realize ultra-wide band gap heterostructures for high frequency device applications.

In summary, we demonstrate silicon delta doping in MOVPE-grown (010)-oriented β-Ga$_2$O$_3$ epitaxial films and β-(Al$_{0.26}$Ga$_{0.74}$)$_2$O$_3$/β-Ga$_2$O$_3$ heterojunctions. Samples A-C show good agreement between SIMS and CV measurements, whereas sample D shows significantly lower delta sheet charge, indicating issues with Si incorporation at high silane flows. We also observed a high concentration of hydrogen in our films at elevated silane flows. HWHM extracted from CV depth profile decreased with increasing sheet charge concentration (6.2 – 3.5 nm). Finally, we report a high electron sheet charge of $6.4 \times 10^{12}$ cm$^{-2}$ by using a delta-doped β-(Al$_{0.26}$Ga$_{0.74}$)$_2$O$_3$ barrier layer. The measured charge could also include a contribution from a parallel channel in the β-(Al$_{0.26}$Ga$_{0.74}$)$_2$O$_3$ alloy barrier. The CV extracted charge density showed no significant change for measurements at 90K and 300 K, confirming the degenerate nature of electron gas. These early results on delta doping show the potential of the MOVPE growth for realization of high density 2DEG at β-(Al$_x$Ga$_{1-x}$)$_2$O$_3$/β-Ga$_2$O$_3$ heterojunction. Further optimization of the growth process is needed to achieve abrupt doping profiles in MOVPE-grown β-Ga$_2$O$_3$.

**Acknowledgments**

This material is based upon work supported by the Air Force Office of Scientific Research under award number FA9550-18-1-0507 monitored by Dr Ali Sayir. Any opinions, finding, and conclusions or recommendations expressed in this material are those of the author and do not necessarily reflect the views of the United States Air Force. This work was performed in part at the Utah Nanofab sponsored by the College of Engineering and the Office of the Vice President for Research. We thank Air Force Research Laboratory, Sensors Directorate for discussions.

## Figure Captions

**Fig. 1.** Charge profile extracted from CV measurements on MOVPE-grown delta-doped β-$Ga_2O_3$ samples A-D. Sample and growth details are listed in Table I.

**Fig. 2.** (a) SIMS stack of MOVPE-grown β-$Ga_2O_3$ epitaxial films with multiple Si delta-sheets (A - G) grown under different silane flows and temperatures (b) SIMS scan of the corresponding stack showing depth profile of Si and H incorporation in the film.

**Fig. 3.** Comparison of measured sheet charge and Si density from CV and SIMS respectively.

**Fig. 4.** (a) Epitaxial structure of modulation-doped β-$(Al_{0.26}Ga_{0.74})_2O_3$/ β-$Ga_2O_3$ heterostructure, (b) HRXRD 2θ-ω scan of the sample, (c) measured CV profile at 90 K and 300 K, and (d) extracted carrier concentration vs depth profile of the electron sheet charge at 90 K and 300 K.

**Table I.** Epitaxial stack, CV and SIMS characterization details of delta-doped β-Ga$_2$O$_3$ thin films used in this study

| Sample | Silane Flow(nmol/min) | Cap Layer Thickness(nm) | Buffer Thickness(nm) | CV Sheet Charge (x $10^{12}$ cm$^{-2}$) | SIMS Si Concentration(x $10^{12}$ atoms.cm$^{-2}$) |
|---|---|---|---|---|---|
| A | 1.9 | 90 | 500 | 2.9 | 3.4 |
| B | 3.8 | 90 | 500 | 6.2 | 6.1 |
| C | 7.7 | 34 | 500 | 9 | 11 |
| D | 17.3 | 20 | 500 | 7.8 | -- |

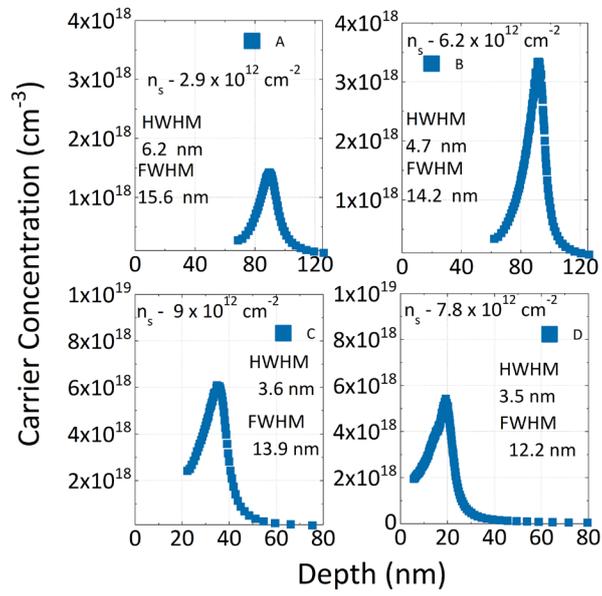

Figure 1

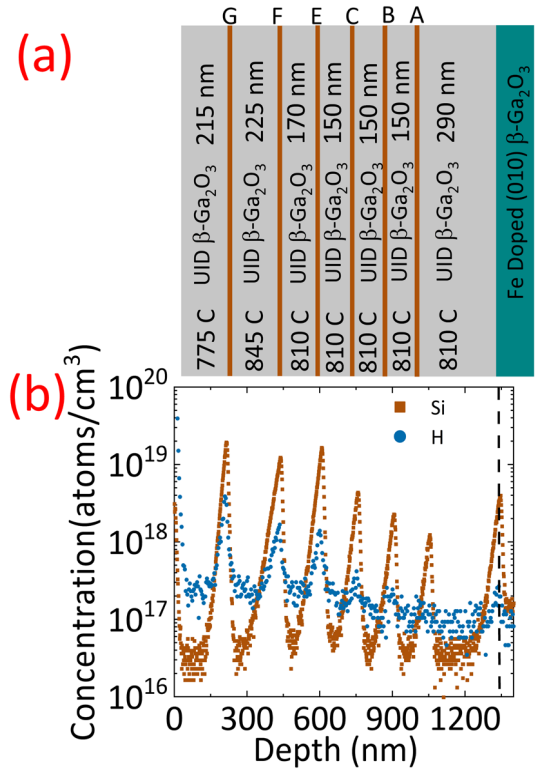

Figure 2

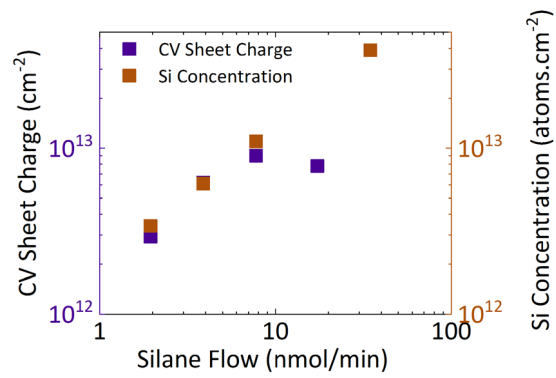

Figure 3

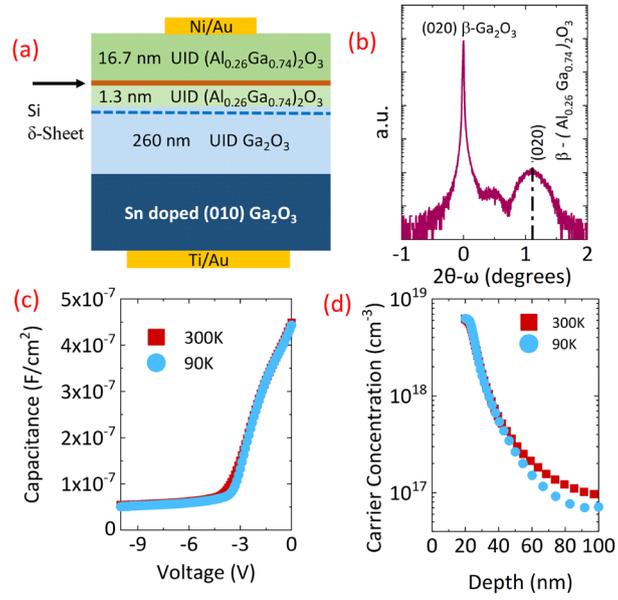

Figure 4

# Delta-doped β-Ga$_2$O$_3$ thin films and β-(Al$_{0.26}$Ga$_{0.74}$)$_2$O$_3$/β-Ga$_2$O$_3$ heterostructures grown by metalorganic vapor-phase epitaxy

## Supplementary data

Praneeth Ranga[1]*, Arkka Bhattacharyya[1], Ashwin Rishinaramangalam[2], Yu Kee Ooi[1], Michael A. Scarpulla[1,3], Daniel Feezell[2], Sriram Krishnamoorthy[1]*

[1] Department of Electrical and Computer Engineering, The University of Utah, Salt Lake City, UT 84112, United States of America

[2] Center for High Technology Materials, University of New Mexico, Albuquerque, NM 87106, United States of America

[3] Department of Materials Science and Engineering, The University of Utah, Salt Lake City, UT, 84112, United States of America

E-mail: praneeth.ranga@utah.edu, sriram.krishnamoorthy@utah.edu

### FWHM comparison of delta-doped β-Ga$_2$O$_3$ thin films

Depth profiles extracted from CV measurements represent the apparent free carrier concentration profiles instead of the donor density distribution. In uniformly doped semiconductors, the resolution of the CV measurement is limited by the Debye length. In order to estimate the FWHM of the delta-doped layers, it is important to make sure that the measured FWHM values are greater than the CV resolution. In degenerate semiconductors with quantum confinement, the CV resolution is not limited by the screening length. Instead, the resolution depends on the spread of the wave function confined by the quantum well (Eq.1). Assuming an ideal delta sheet of donors, the wave function spread ($\Delta z_{cv}$) can be approximated as[22)],

$$\Delta z_{cv} = 2\left(\frac{7}{5}\right)^{0.5}\left(\frac{4\varepsilon\hbar^2}{9e^2 N_{2D} m^*}\right)^{(1/3)} \qquad (1)$$

Where ℏ is the reduced Planck's constant, ε is the dielectric constant of (010) oriented β-$Ga_2O_3$(10), $m^*$ is the conduction band effective mass in β-$Ga_2O_3$(0.28 $m_0$), e is charge of the electron and $N_{2D}$ is the electron sheet charge density. Figure. S1 shows the theoretical FWHM for an ideal delta sheet and FWHM measured from CV and SIMS measurements. Clearly, the FWHM measured by CV and the Si FWHM measured using SIMS is much higher than the ideal delta doping profile, indicating that our measurements are not limited by the CV resolution. In Fig. S1 we also observe minimal difference between FWHM values of samples with different silane flow, indicating that the silane flow doesn't have a significant effect on SIMS FWHM. The FWHM measured using SIMS profile reduces with decreasing growth temperature. Layer G grown at a temperature of 775 °C has a FWHM of 16 nm compared to 18 nm and 23 nm of layers E (810 °C) and F (845 °C), respectively. This indicates that the spread of Si dopants is strongly determined by growth temperature. It should be noted that the FWHM of delta-doped layer grown using molecular beam epitaxy is rather close to the expected wave function spread[23].

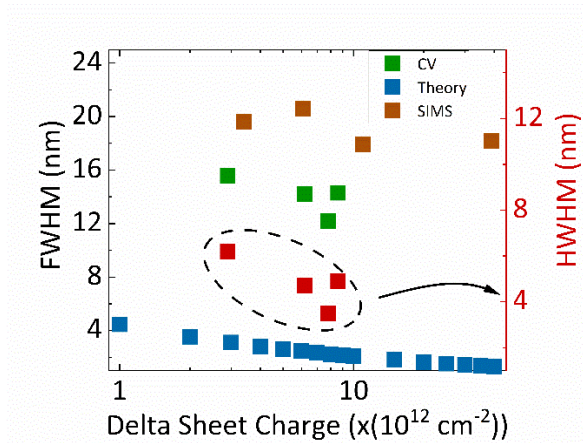

**Fig. S1**

Fig. S1  Plot of theoretical FWHM along with FWHM and HWHM measured from SIMS and CV characterization.

# Band diagram of delta-doped β-(Al$_{0.26}$Ga$_{0.74}$)$_2$O$_3$/β-Ga$_2$O$_3$ heterojunction

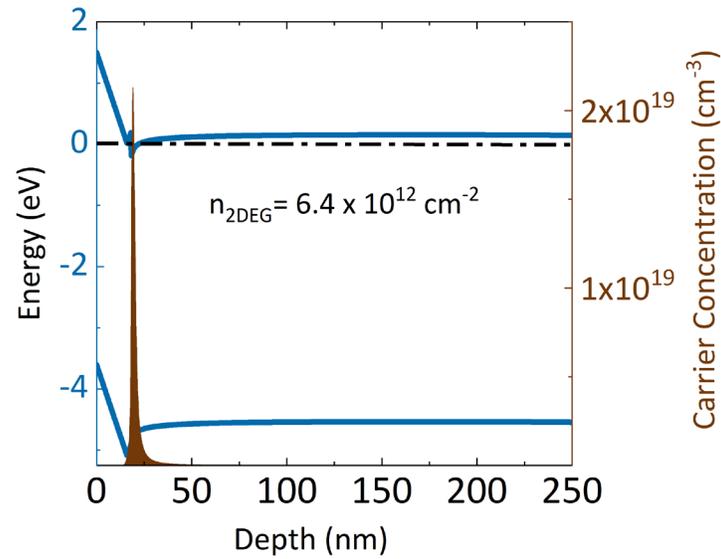

**Fig.S2**

Fig.S2 shows the calculated band diagram of β-(Al$_{0.26}$Ga$_{0.74}$)$_2$O$_3$/β-Ga$_2$O$_3$ heterojunction in Fig.4(a). Assuming a delta sheet of 1.1 x 10$^{13}$ cm$^{-2}$ with complete ionization, band offset of 0.4 eV and a barrier height of 1.5 eV, we estimate a total electron sheet charge of 6.4 x 10$^{12}$ cm$^{-2}$.

## AFM scan of β-(Al$_{0.26}$Ga$_{0.74}$)$_2$O$_3$/β-Ga$_2$O$_3$ heterojunction

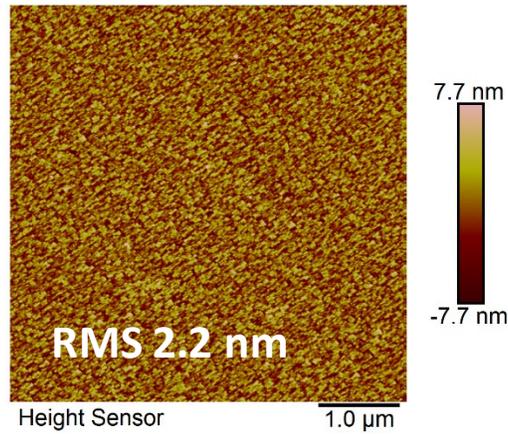

**Fig. S3**

Fig.S3 shows a 5 μm x 5 μm scan of β-(Al$_{0.26}$Ga$_{0.74}$)$_2$O$_3$/β-Ga$_2$O$_3$ heterojunction in Fig.4(a). The RMS roughness of the AFM scan is 2.2 nms.

# Delta-doped β-(Al$_{0.26}$Ga$_{0.74}$)$_2$O$_3$/β-Ga$_2$O$_3$ heterojunction with a thick spacer layer

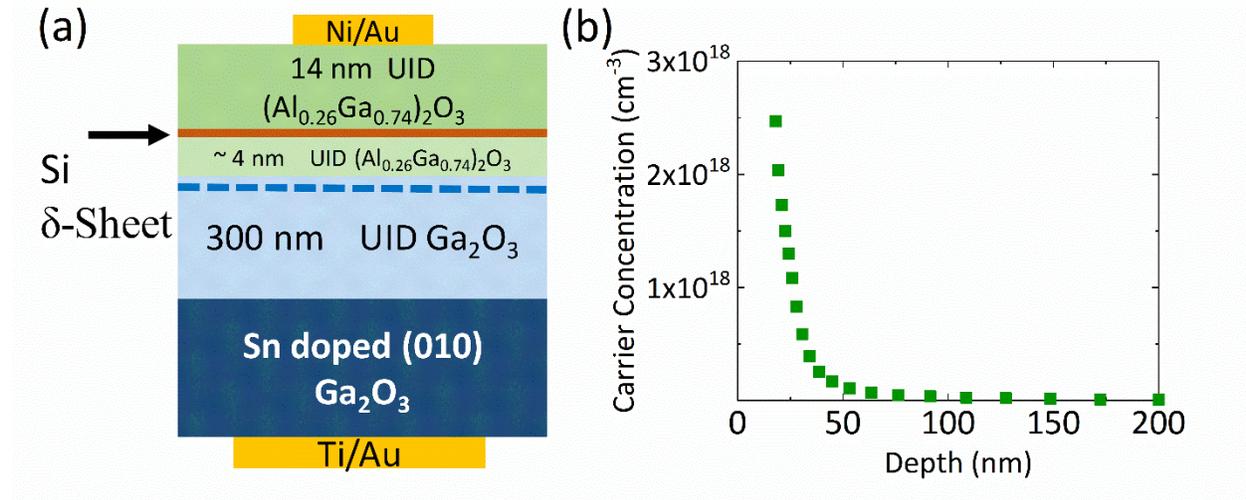

**Fig.S4**

To study modulation doping in delta-doped β-(Al$_{0.26}$Ga$_{0.74}$)$_2$O$_3$/β-Ga$_2$O$_3$ heterojunction, we increased the β-(Al$_{0.26}$Ga$_{0.74}$)$_2$O$_3$ spacer thickness to 4 nm and kept the silane flow and total β-(Al$_{0.26}$Ga$_{0.74}$)$_2$O$_3$ thickness identical to the heterostructure in Fig.4. We observed that the CV measured 2DEG sheet charge reduced to 2.5 x 10$^{12}$ cm$^{-2}$, which is expected from reduction in modulation doping efficiency (Fig.S4). A thicker spacer layer will increase the probability of parallel channel formation in the β-(Al$_{0.26}$Ga$_{0.74}$)$_2$O$_3$ layer. However, if a parallel channel is present, the peak of the charge profile would be located in the AlGaO layer (<18 nm) at equilibrium.

## Forward bias CV profile of delta-doped β-(Al$_{0.26}$Ga$_{0.74}$)$_2$O$_3$/β-Ga$_2$O$_3$ heterojunction

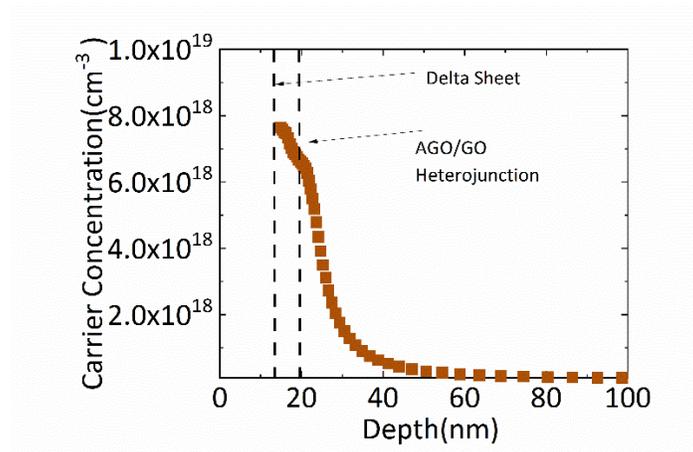

**Fig. S5**

Employing forward bias CV measurements, we are able to profile electron concentration in the β-(Al$_{0.26}$Ga$_{0.74}$)$_2$O$_3$ barrier layer shown in Fig.4(a). The measured charge profile clearly shows a peak close to 15 nm (Fig.S5), indicating that the silicon delta sheet is located in the β-(Al$_{0.26}$Ga$_{0.74}$)$_2$O$_3$ barrier layer.